\documentclass[11pt,twoside]{article}

\usepackage{asp2006}
\usepackage{epsf}
\usepackage{psfig}
\usepackage{lscape}

\begin{document}
\title{The Ages of Early--Type Galaxies at $z\sim 1$}   
\author{Sperello di Serego Alighieri}   
\affil{INAF -- Osservatorio Astrofisico di Arcetri, Largo E. Fermi 5,
50125 Firenze, Italy}    

\author{Alessandro Bressan}
\affil{INAF -- Osservatorio Astronomico di Padova, Vicolo
dell'Osservatorio 5, 35100 Padova, Italy}

\author{Lucia Pozzetti}
\affil{INAF -- Osservatorio Astronomico di Bologna, Via Ranzani 1, 40127
Bologna, Italy}

\begin{abstract} 
The study of the ages of early--type galaxies and their dependence on
galaxy mass and environment is crucial for understanding the formation
and early evolution of galaxies. We review recent works on the ${\cal
M}/L$ ratio evolution, as derived from an analysis of the Fundamental 
Plane of early--type galaxies at $z\sim 1$ both in the field and in the 
clusters environment. We use the ${\cal M}/L$ ratio to derive an estimate 
of the galaxy age. We also use a set of high--S/N intermediate--resolution 
VLT spectra of a sample of early--type galaxies with $0.88<z<1.3$ from 
the K20 survey to derive an independent estimate of their age by fitting 
SSP model spectra. Taking advantage of the good leverage provided by 
the high sample redshift, we analyse the results in comparison with the 
ages obtained for the same sample from the analysis of the ${\cal M}/L$
ratio, and with the predictions of the current hierarchical models of 
galaxy formation.
\end{abstract}

\section{Introduction}   

Everyone knows that the year of birth can be determined at first sight
more accurately for a child rather than for a middle--aged person (except
if he is celebrating his 60th anniversary!). The same is
true for the ages of galaxies. Furthermore the finite speed of light and 
the large size of the Universe give us the possibility to study the 
galaxies which are now old, when they were a lot younger, by observing
them at high redshift. This is
precisely what we are trying to do with early--type galaxies (ETG), which
are particularly important to date, since they contain most of the visible
mass in the Universe and they trace its highest density peaks
\citep{ren06}.
\begin{figure}[!ht]
\plotone{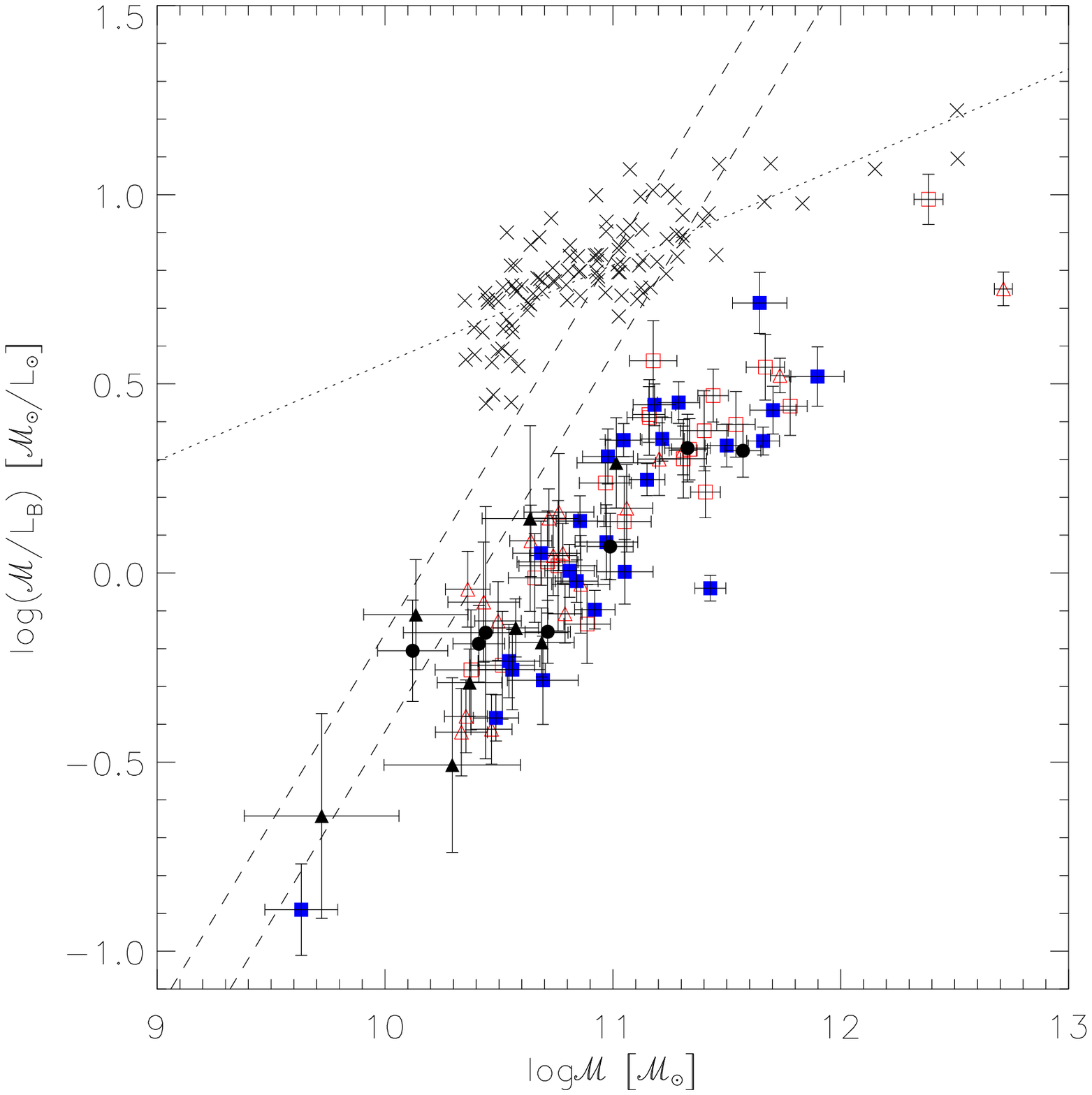}
\caption{The ${\cal M}/L$ ratio in the B-band as a function of galaxy mass
for local ETG in the Coma
Cluster \citep{jor06} (black crosses), for field ETG at $z$$\sim$1 from
the
K20 survey \citep{dsa05} both in the CDFS field (filled black circles)
and in the Q0055 field (filled black triangles), for field ETG at
$z$$\sim$1
in the GOODS area \citep{tre05} (filled blue squares), and for the ETG in
two
clusters \citep{jor06} at z=0.835 (open red squares) and at z=0.892 (open
red
triangles). The dotted line is
a fit to the Coma ETG, while the upper and lower dashed lines represent
the
$M_B=-20.0$ and
$M_B=-20.5$
magnitude limits of di Serego Alighieri et al. (2005) and of J\o rgensen
et
al. (2006) respectively. The changes in ${\cal M}/L_B$ from
high redshift to $z=0$ decrease with galaxy mass in all environments and
are
similar in the field and in the clusters.}
\end{figure}
ETG are very regular structures, in fact a two--parameter family,
as testified by the existence and tightness of the Fundamental Plane (FP, 
Djorgovski \& Davis 1987). We can then hope that a relatively simple
process has imprinted their characteristics during formation and evolution.
A straightforward way to find the parameters driving this process is to
look for correlations with the galaxy ages.

\section{Ages from the analysis of the fundamental plane}

The analysis of the FP provides a simple way to estimate galaxy ages
\citep{dsa06a}.
The FP can be seen as a relationship between the ${\cal M}/L$ ratio and the mass,
by a simple coordinate conversion \citep{ben92}. Figure 1 shows a
compilation in these coordinates of the FP data on ETG at $z\sim 1$ both
in the field \citep{dsa05,tre05} and in the clusters \citep{jor06},
compared to the ETG at $z=0$.
The evolution of the ${\cal M}/L$ ratio in the last
10 Gyr is larger for the smaller mass galaxies. Although this trend is
enlarged by selection effects due to the magnitude limited samples, as
shown by the dashed lines in Figure 1, nevertheless it cannot be totally
explained by them \citep{vdw05}. 

The usual way to analyse the evolution in ${\cal M}/L$ ratio as a function of
redshift is to compare it with the fit done by \citet{van03} for the
massive ETG (${\cal M}>10^{11}{\cal M}_{\sun}$) of several clusters for 
redshifts up to z=1.3. However this procedure is largely unsatisfactory, 
since, first, these massive cluster galaxies are not necesarily a uniform
reference class, and, second, it prevents by construction from studying
the lower mass cluster galaxies, which are those more likely to show any
downsizing effect. Therefore, in order to study how the star formation
history of ETG depends on the galaxy mass and the environment, we are
proposing a different approach \citep{dsa06a}, which consists in
interpreting the changes in ${\cal M}/L$ ratio as age differences. In fact it has
been shown that 
other possible interpretations, i.e. systematic structural
changes and partial support by rotation, can only explain a small fraction
of the
observed differential evolution of ${\cal M}/L_B$, and that this evolution
correlates with the rest-frame $U-B$ colour, thereby providing
independent evidence for changes in the stellar populations
\citep{dsa05}. \citet{mar05} has used evolutionary population synthesis
models to estimate the ${\cal M}/L$ for simple stellar populations with
variable age (Fig. 2).
\begin{figure}[!ht]
\plotone{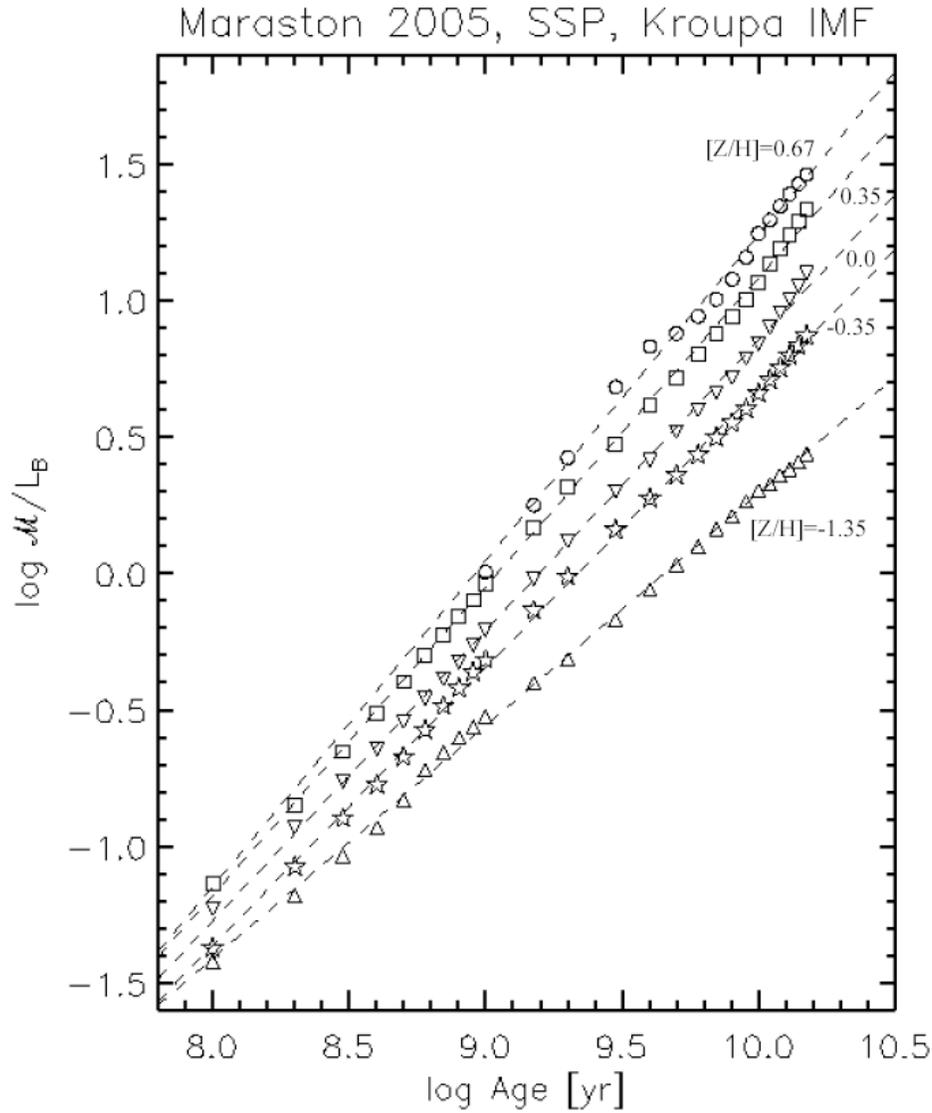}
\caption{The dependence of the ${\cal M}/L_B$ ratio on the age as derived
by \citet{mar05} for simple stellar populations with a \citet{kro01} IMF
for five different metallicities.}
\end{figure}
We have then estimated the luminosity weighted average stellar age for 
each $z\sim 1$ ETG from the ${\cal M}/L_B$ obtained from the FP parameters 
and using the metallicity derived from the observed velocity dispersion 
\citep{tho05}. The results are shown in Figure 3 and clearly show the
downsizing effect, i.e. the age correlates with the galaxy mass both in the
clusters and in the field, and there appears to be no difference in age
due to the environment \citep{dsa06a,dsa06b}.
\begin{figure}[!ht]
\plotone{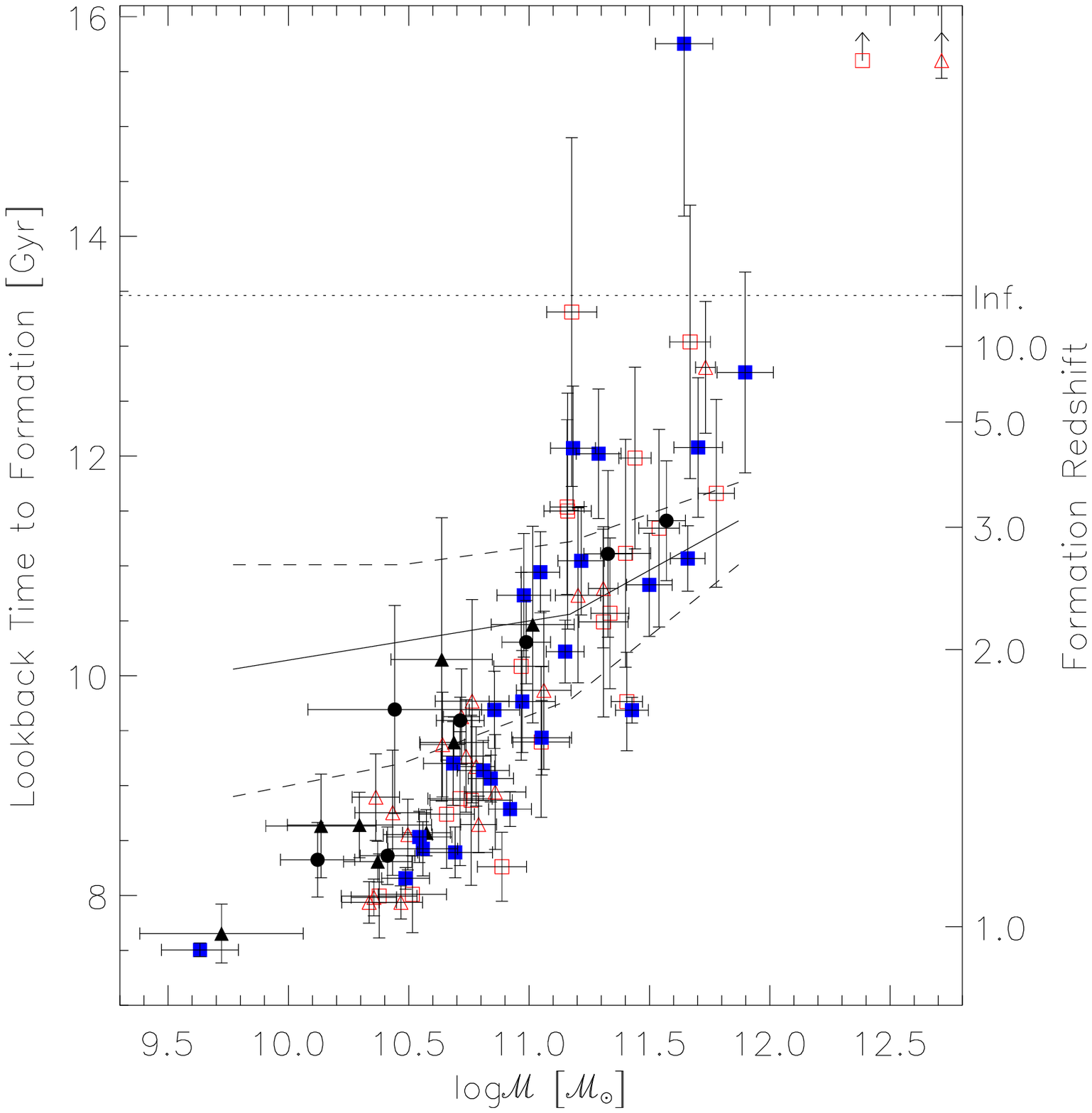}
\caption{The formation epoch for the ETG shown in Figure 1
(same symbols), evaluated as explained in the text. The two upward
pointing
arrows indicate that the two most massive cluster ETG are out of the
figure
(their ages amount to 16.4 and 23.4 Gyr). The continuous line
shows the median model ages obtained by \citet{del06} from a semianalytic
model of hierarchical galaxy evolution, while the dashed lines are their
upper and lower quartiles.
More massive ETG form earlier in all environments, and the ages are not
influenced by the environment.}
\end{figure}
Since for a given ${\cal M}/L$ ratio the age decreases with metallicity 
(Fig. 2) and velocity dispersion, hence with mass, the downsizing effect 
would be even stronger, if we assume a fixed metallicity for all galaxies.

\section{Ages from fits of spectral synthesis model to the spectra}

Looking for an independent estimate of the ages obtained from the FP
para\-meters, we have recently been experimenting with ages obtained by fitting the
observed intermediate resolution spectra of the $z\sim 1$ field ETG from
the K20 survey \citep{dsa05} with synthetic model spectra (Pozzetti et al.
in preparation). Preliminary results have been obtained using high resolution
SSP--based atmosphere models by Bertone et al. (in preparation) and the
metallicitiy derived from the velocity dispersion, as in the previous
section. Figure 4 shows the comparison of these preliminary results with
the ages obtained from the FP parameters. This comparison shows that,
although for the majority of the galaxies the ages obtained with the two
methods are consistent, there are a number of cases which are discrepant,
some with larger FP ages and others with smaller ones. More work needs
therefore to be done to understand these discrepancies and to gain
sufficient confidence in the estimated ETG ages. For this purpose we are
also planning to constrain ages using the spectral indices (e.g. Renzini
2006 and references therein).

\begin{figure}[!ht]
\plotone{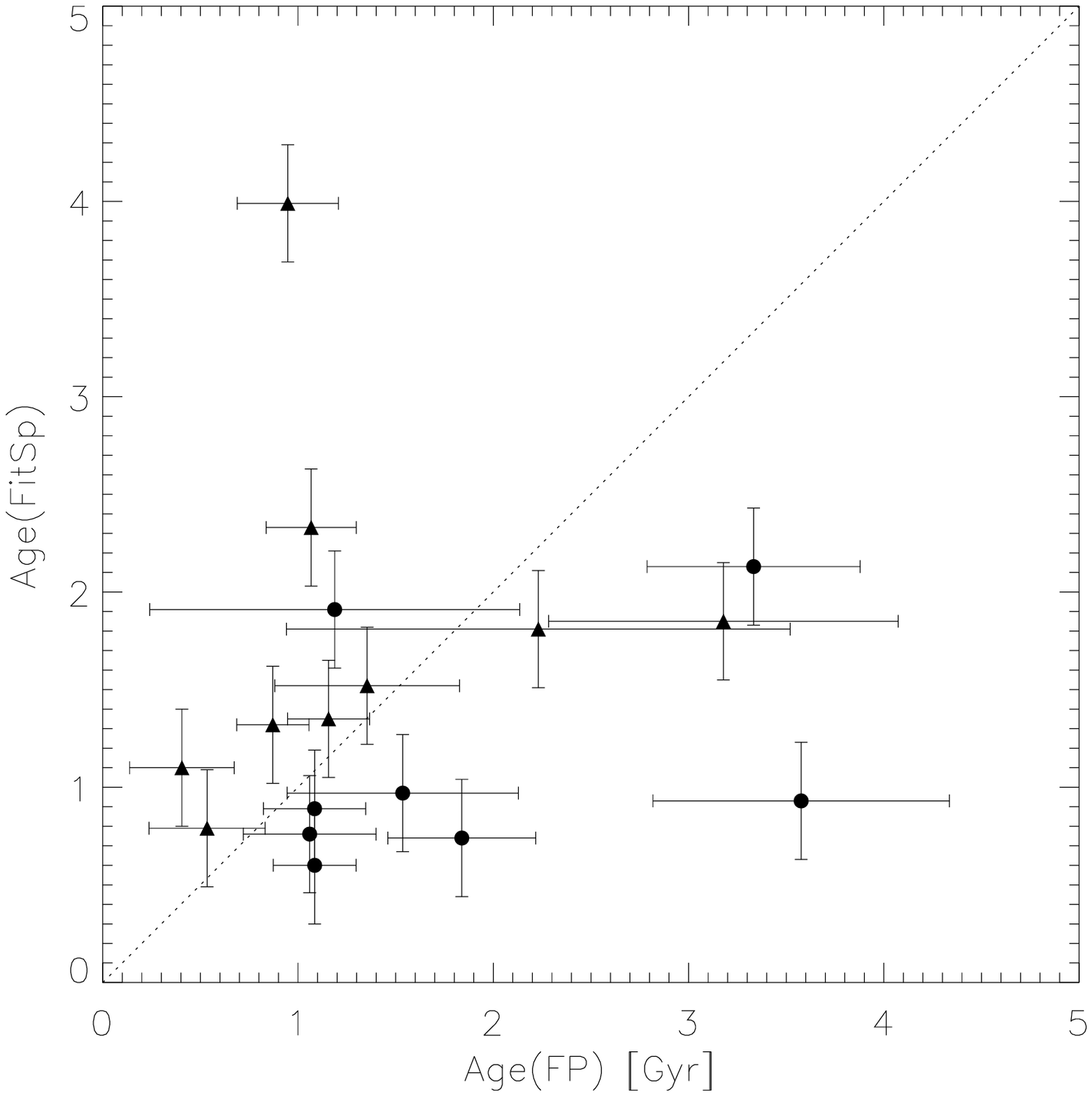}
\caption{Comparison of the ages obtained from the FP parameters with those
obtained from fits of spectral synthesis models to the spectra of ETG at 
$z\sim 1$ from the K20 survey  of \citet{dsa05}.}
\end{figure}

\section{Summary and future work}

We have used the scaling parameters of ETG (luminosity, effective radius,
and velocity dispersion) to evaluate ${\cal M}/L_B$  ratio changes as a
function of redshift (for $0<z<1.3$), galaxy mass and environment. The
${\cal M}/L_B$ ratio evolves faster for lower mass galaxies both in the
field and in the clusters, and the evolution does not depend on the
environment. Interpreting the ${\cal M}/L$ ratio changes as age
differences we infer that the age increases with mass (downsizing) both in
the field and in the clusters, and field galaxies have the same age as
cluster galaxies with the same mass. We are currently obtaining independent
estimates of the age of the $z\sim 1$ ETG by fitting the observed spectra
with sythesis models and by using the spectral line indices. In addition we 
will use a different independent approach by deriving the metallicity (and
the related age) directly from the fits to the spectra.

\end{document}